\documentclass[aps,amsmath,amssymb,amsfonts,preprint,nofootinbib]{revtex4}

\usepackage{graphicx}

\usepackage{tikz}
\usepackage{caption}
\usepackage{subcaption}
\usepackage{mathtools}

\usetikzlibrary{shapes.misc}
\usetikzlibrary{arrows.meta}
\tikzset{cross/.style={cross out, draw=black, minimum size=2*(#1-\pgflinewidth), inner sep=0pt, outer sep=0pt},
cross/.default={1pt}}
\tikzstyle{axisarrow} = [-{Latex[inset=0pt,length=5pt]}]

\newcommand{\ZZ}{{\mathbb Z}}
\newcommand{\RR}{{\mathbb R}}
\newcommand{\ra}{\rightarrow}

\newcommand{\Tr}{{\rm Tr}}

\newcommand{\cf}{{\mathcal F}}

\usepackage{amsmath}

\newcommand{\cB}{{\mathcal B}}
\newcommand{\cL}{{\mathcal L}}
\newcommand{\cM}{{\mathsf M}}
\newcommand{\fM}{{\mathfrak M}}
\newcommand{\C}{{\mathcal C}}

\newcommand{\cm}{{\mathbf m}}
\newcommand{\cs}{{\mathbf s}}

\newcommand{\Fth}{F^{(3)}}

\begin{document}

\title{Higher-dimensional generalizations of the Berry curvature}
\author{Anton Kapustin}
\email{kapustin@theory.caltech.edu}
\author{Lev Spodyneiko}
\email{lionspo@caltech.edu}
\affiliation{California Institute of Technology, Pasadena, CA 91125, United States}

\begin{abstract}
A family of finite-dimensional quantum systems with a non-degenerate ground state gives rise to a closed 2-form on the parameter space: the curvature of the Berry connection. Its cohomology class is a topological invariant of the family. We seek generalizations of the Berry curvature to families of gapped many-body systems in $D$ spatial dimensions. Field theory predicts that in spatial dimension $D$ the analog of the Berry curvature is a closed $(D+2)$-form on the parameter space (the Wess-Zumino-Witten form). We construct such closed forms for arbitrary families of interacting lattice systems in all dimensions. In the special case of systems of free fermions in one dimension, we show that these forms can be expressed in terms of the Bloch-Berry connection on the product of the Brillouin zone and the parameter space. In the case of families of Short-Range Entangled systems, we argue that integrals of our forms over spherical cycles are quantized. 
\end{abstract}

\maketitle

\section{Introduction}

Consider a quantum-mechanical system with a Hamiltonian depending on parameters, a unique ground state for all values of the parameters, and an energy gap to the lowest excited state. To these data one can associate a 2-form $\Omega$ on the parameter space called the Berry curvature \cite{Berry}. This 2-form is closed and quantized:  its periods (integrals over closed surfaces in the parameter space $\cM$) are integral multiples of $2\pi$. Quantization of the periods of the Berry curvature can be explained as follows. On the one hand, the ground states of the system fit into a rank-one complex vector bundle $\cB$ over $\cM$. The Berry curvature $\Omega$ is the curvature of a certain unitary connection on this vector bundle (the Berry connection). On the other hand, it is well-known that {\it any} rank-one vector bundle $\cL$ over a manifold $\cM$ has a topological invariant called the 1st Chern class taking values in $H^2(\cM,\ZZ)$. Its image in the de Rham cohomology $H^2(\cM,\RR)$ can be represented by $F/2\pi$, where $F$ is the curvature 2-form of {\it any} unitary  connection on $\cL$. These two facts imply that periods of $\Omega$ are $2\pi$ times an integer. 


While definitions of the Berry connection and the Berry curvature do not make an explicit reference to the dimension of space, they are not directly applicable to models of quantum statistical mechanics in spatial dimension $D\geq 1$, or to models of Quantum Field Theory. The reason is that the Hilbert space of these models  is infinite-dimensional, and consequently the formulas defining the Berry connection and Berry curvature may be afflicted with divergences. In the case of models of statistical mechanics these are volume divergences, while in the case of QFT there can be both volume and short-distance divergences. The presence of volume divergences is easy to see in the case of an infinite system of identical decoupled spins coupled to a magnetic field. Each spin contributes additively to the Berry curvature 2-form, therefore the total Berry curvature is proportional to the volume of the system. One could try to define Berry curvature per unit volume, but its existence and quantization for more general extended systems is far from clear. 

Even if the Berry curvature could somehow be defined for infinitely extended systems, its physical and mathematical interpretation would be unclear. For such systems, only the algebra of observables is independent of the parameters of the Hamiltonian. The Hilbert space is not specified from the outset. It depends on the choice of a suitable state on the algebra of observables via the Gelfand-Naimark-Segal construction. This state is a ground state of the Hamiltonian and thus itself depends on the parameters. There is no natural way to identify Hilbert spaces for different values of the parameters, and thus ground states do not form a well-defined vector bundle. Therefore one would not be able to interpret the Berry curvature as the curvature of a connection on a rank-one vector bundle.

A few years ago A. Kitaev proposed that for a family of Short-Range Entangled (SRE) gapped systems in spatial dimensions $D$ one can define a closed $(D+2)$-form on the parameter space. This form is a higher-dimensional generalization of the Berry curvature. The cohomology class of this $(D+2)$-form would serve as a topological invariant of the family. One difficulty in making this proposal concrete is that currently there is no useful definition of Short-Range Entangled systems, beyond the "negative" statement that these are systems which exhibit neither spontaneous symmetry-breaking nor topological order.

Unlike the notion of a Short-Range Entangled system, the notion of a gapped system is straightforward to define. 
In this paper we define and study higher-dimensional generalizations of the Berry curvature for gapped lattice systems on $\RR^D$. For  any family of such systems we define a closed $(D+2)$-form $\Omega^{(D+2)}$ on the parameter space $\cM$. The form depends on some choices (essentially, a choice of a differential $D$-form on $\RR^D$ whose integral over $\RR^D$ is $1$). The cohomology class of $\Omega^{(D+2)}$ is independent of any choices and is a topological invariant of the family. It is an obstruction to continuously deforming the family to a constant family of gapped systems. It also can be viewed as an obstruction to having a gapped boundary which varies continuously with parameters. We also argue that when all systems in the family are Short-Range Entangled, the integral of $\Omega^{(D+2)}$ over any $(D+2)$-dimensional sphere is an integral multiple of $2\pi$. In this special case our set-up matches that in Kitaev's proposal. 

For families of Euclidean lattice systems in $D+1$ dimensions with exponentially decaying correlations, A. Kitaev outlined a construction of a closed $(D+2)$-form on the parameter space \cite{Kitaev_talk}. Our results can be viewed as a Hamiltonian version of this construction.

In the case of translationally-invariant tight-binding free fermion Hamiltonians in 1d we show that the cohomology class of $\Omega^{(D+2)}$ is determined by the curvature of the Berry-Bloch connection. We conjecture that this is true in any dimension. Free fermion systems provide examples of families whose topological invariants are non-trivial. 

Recently Cordova, Freed, Lam, and Seiberg studied field theories with "anomalies in the space of couplings" \cite{CFLS1,CFLS2}. Via the bulk-boundary correspondence, this subject is closely related to topologically-nontrivial families of gapped field theories in one dimension higher. It is natural to conjecture that there is a 1-1 correspondence between topological invariants of families of gapped field theories in $(D+1)$ space-time dimensions and topological invariants of families of gapped lattice models in $D$ spatial dimensions some of which we study here.

The content of the paper as follows. In Section \ref{sec:effective action} we interpret higher Berry curvature forms in the language of Quantum  Field Theory, specifically as Wess-Zumino-Witten terms in the effective action for the parameters. This serves as a motivation for subsequent discussion. In Section \ref{sec:one dimension} we show how to associate a closed 3-form to a family of gapped 1d lattice systems. In Section \ref{sec:arbitrary dimension} we extend our construction to families of gapped lattice systems in arbitrary spatial dimension. This requires some mathematical machinery which we review. We discuss our results in Section \ref{sec:discussion}. In Appendix A we argue that if all systems in the family are Short-Range Entangled, the integral of 
$\Omega^{(D+2)}$ over any spherical cycle in the parameter space is an integral multiple of $2\pi$. We also explain the interpretation of the period of $\Omega^{(D+2)}$ on $S^{D+2}$ as an obstruction to having a gapped boundary condition defined globally on $S^{D+2}$. In Appendix B we compute the 3-form $\Omega^{(3)}$ for families of tight-binding free fermion 1d systems of class A and express its cohomology class in terms of the Berry-Bloch connection. This allows us to give examples of families of systems where our higher Berry curvature is topologically non-trivial (lies in nonzero cohomology classes).

 A. K. would like to thank the other members of the gang of the seven (D. Freed, M. Freedman, M. Hopkins, A. Kitaev. G. Moore, and C. Teleman) for discussions of family invariants of gapped systems and related issues. We are especially grateful to A. Kitaev for reading a preliminary draft of the paper and pointing out an error. This research was supported in part by the U.S.\ Department of Energy, Office of Science, Office of High Energy Physics, under Award Number DE-SC0011632. A.K. was also supported by the Simons Investigator Award.

\section{Effective action considerations}\label{sec:effective action}

The purpose of this section is to motivate the constructions in subsequent sections. It is not essential for understanding the rest of the paper. Readers not familiar with topological aspects of QFT are advised to skip it on first reading.

If a gapped system in $D$ spatial dimensions is described by a trivial topological field theory at long distances, then its low-energy effective action is a well-defined function of background fields, such as the metric and the gauge fields which couple to global symmetries. If one deals with a family of such systems parameterized by a manifold $\cM$, one can let the parameters vary slowly from point to point, and the effective action is still a well-defined function of the background fields. The variation of the parameters can be described by a map $\phi: X\ra \cM$, where $X$ is the space-time. The effective action depends on $\phi$ as well as other background fields.

Loosely speaking, topological terms in the action are those terms which survive when one re-scales the metric $g_{\mu\nu}\mapsto e^{\sigma} g_{\mu\nu}$ and takes the limit $\sigma\ra +\infty$. The simplest such terms are those which depend only on $\phi$ and not on other background fields. For example, for $D=0$ (ordinary quantum mechanics) such a topological term schematically has the form
\begin{equation}
S_{top}(X,\phi)=\int_X \omega^{(1)}_j \partial_t\phi^j dt=\int_X\phi^*\left(\omega^{(1)}\right),
\end{equation}
where $X$ is a one-dimensional manifold ($S^1$ or $\RR$) and $\omega^{(1)}$ is the 1-form on $\cM$ representing the Berry connection. This formula is only schematic because in general the Berry connection on the parameter space can be represented by a 1-form $\omega^{(1)}$ only locally on $\cM$. If the cohomology class of the Berry curvature $\Omega^{(2)}$ is non-trivial, then one cannot write $\Omega^{(2)}=d\omega^{(1)}$ for a globally-defined 1-form $\omega^{(1)}$. Rather, one needs to cover the parameter space with charts, in each of which the connection is represented by a 1-form. On the overlaps of the charts these 1-forms are related by gauge transformations. To define $S_{top}(X,\phi)$ properly, one needs to know both the locally-defined 1-forms and the gauge transformations connecting them. 

Another important point is that only $\exp(i S_{top}(X,\phi))$ can be defined unambiguously, while $S_{top}(X,\phi)$ is defined only up to an integer multiple of $2\pi$. To see this, let us pick an oriented two-dimensional manifold $Y$ such that $\partial Y=X$. If the map $\phi:X\ra \cM$ extends to a continuous map $\tilde\phi: Y\ra \cM,$ then one can write a more precise formula for the topological action as follows:
\begin{equation}
S_{top}=\int_Y \tilde\phi^*\left(\Omega^{(2)}\right).
\end{equation}
This expression depends on the choice of $\tilde\phi$. But $\exp(iS_{top}(X,\phi))$ is unambiguously defined since periods of $\Omega^{(2)}$ are "quantized": the integral of $\Omega^{(2)}$ over any 2-cycle on $\cM$ is $2\pi$ times an integer. If $\phi$ does not extend to $Y$, one can still define $\exp(iS_{top}(X,\phi))$ as the holonomy of the Berry connection, but to compute it one needs to use choose local trivializations, as sketched above.

Although the cohomology class of $\Omega^{(2)}$ does not completely determine the Berry connection, it does determine it up to an addition of a globally defined 1-form. Since any 1-form can be deformed to zero, this means that the cohomology class of the Berry curvature determines $\exp(iS_{top}(X,\phi))$ up to a continuous deformation. This cohomology class is often easier to compute than the Berry curvature itself, because the Berry curvature is not a deformation invariant and depends on dynamical details.

For $D>0$ the story is similar. A topological action which does not depend on fields other than $\phi$ schematically has the form
\begin{equation}\label{WZWd}
S_{top}(X,\phi)=\int_X \phi^*\left(\omega^{(D+1)}\right)=\frac{1}{(D+1)!}\int_X \omega^{(D+1)}_{i_0\ldots i_{D}} \left(\partial_0 \phi^{i_0}\right)\ldots\left(\partial_{D}\phi^{i_{D}}\right)dx^0\ldots dx^{D},
\end{equation}
where $\omega^{(D+1)}$ is an $(D+1)$-form on $\cM$ and $X$ is a closed oriented $(D+1)$-manifold. If one takes this formula literally, then all such actions can be deformed to zero, since any $(D+1)$-form can be deformed to zero. But if one interprets $\omega^{(D+1)}$ more creatively, as a sort of "higher connection", one can get more interesting actions which cannot be deformed to the trivial one. One way to find such a generalization is to note that the r.h.s. of the above equation does not change under $\omega^{(D+1)}\mapsto\omega^{(D+1)}+d\lambda^{(D)}$, where $\lambda^{(D)}$ is an arbitrary $D$-form. Then it is natural to consider an object specified by locally-defined $(D+1)$-forms $\omega^{(D+1)}_\alpha$, where $\alpha$ labels the charts. On the overlaps of charts these $(D+1)$-forms are related by $D$-form gauge transformations. The full story is rather complicated, since in order to be able to define "higher holonomy" along a $(D+1)$-dimensional submanifold one needs  compatibility conditions for the gauge transformations which involve $(D-1)$-forms on triple overlaps, etc. 

An alternative approach (first appearing in a mathematical paper by Cheeger and Simons \cite{CheegerSimons}) is to postulate the following natural property. If $X=\partial Y$ for some $(D+2)$-manifold $Y$, and if $\phi$ extends to a map $\tilde\phi:Y\ra \cM$, then one must have
\begin{equation}
\exp\left(i S_{top}(X,\phi)\right)=\exp\left(i\int_Y\tilde\phi^*\left(\Omega^{(D+2)}\right)\right),
\end{equation} 
where $\Omega^{(D+2)}$ is a $(D+2)$-form on $\cM$. For this formula to make sense, $\Omega^{(D+2)}$ must be closed and its periods must be integer multiples of $2\pi$. For example, to see that $\Omega^{(D+2)}$ must be closed, one can vary $\tilde\phi$ infinitesimally while keeping its boundary value $\phi$ fixed. It is easy to see that the r.h.s. will be unchanged only if $d\Omega^{(D+2)}=0$. To see that $\Omega^{(D+2)}$ must have periods which are integral multiples of $2\pi$, take $X$ to be the empty manifold, and take $Y$ to be any closed $(D+2)$-manifold.

Locally on $\cM$ one can write $\Omega^{(D+2)}=d\omega^{(D+1)}$. If the cohomology class of $\Omega^{(D+2)}$ is trivial, one can do it globally, and then $S_{top}(X,\phi)$ can be defined by the simple formula (\ref{WZWd}). In general, one can show that given a closed $(D+2)$-form $\Omega^{(D+2)}$ with "quantized" periods there exists an exponentiated action $\exp(i S_{top}(X,\phi))$ satisfying the above equation. It is unique up to a factor $\exp(i\int_X\phi^*(\alpha))$, where $\alpha$ is a closed $(D+1)$-form on $\cM$. 

As in the case $D=0$, this implies that the cohomology class of $\Omega^{(D+2)}$ determines $\exp(i S_{top}(X,\phi))$ up to a factor which can be deformed to $1$. Thus one can say that deformation classes of such topological actions (known as Wess-Zumino-Witten terms) are classified by "quantized" cohomology classes of degree $D+2.$ There is also an interpretation of Wess-Zumino-Witten terms as  holonomies of "higher connections" on "higher bundles" on $\cM$. Then the cohomology class of $\Omega^{(D+2)}$ determines the topology of the corresponding "higher bundle". But since such an interpretation is quite abstract, we will not use it in this paper.

The conclusion is that given a family of trivial gapped systems in spatial dimension $D$, one should be able to obtain a closed $(D+2)$-form on the parameter space with "quantized" periods. While the form itself depends on the dynamical details, its cohomology class is a topological  invariant. It classifies possible deformation classes of Wess-Zumino-Witten terms on the parameter space. 

The statement about quantization of periods needs some qualification in the case of fermionic systems. A fermionic path-integral depends on spin structure on $X$. For fermionic systems it is unreasonable to restrict attention to topological terms which depend only on the map $\phi$, one needs to study topological terms which depend both on $\phi$ and the spin structure. Then one needs to generalize the Cheeger-Simons approach by requiring the manifolds $X$ and $Y$ to be spin manifolds. Such spin-structure-dependent Wess-Zumino-Witten terms were first considered in \cite{Freed pions}. Alternatively, if one limits oneself to the case of systems on $X=\RR^{D+1}$ or its one-point compactification $S^{D+1}$, then one can always take $Y=B^{D+2}$ ($(D+2)$-dimensional ball). Then the quantization condition is relaxed: only integrals of the form
\begin{equation}
\int_{S^{D+2}} h^*\left( \Omega^{(D+2)}\right)
\end{equation}
need to be integral multiples of $2\pi$. Here $h:S^{D+2}\ra \cM$ is any smooth map. We will call such an $h$ a spherical cycle. Thus for fermionic systems only integrals of $\Omega^{(D+2)}$ over spherical cycles are quantized. Of course, not all topological terms which are consistent on $\RR^{D+1}$ or $S^{D+1}$ will remain consistent when considered on a general space-time. That is, quantization on spherical cycles is not enough to make the Wess-Zumino-Witten action well-defined on arbitrary spin manifolds.

\section{Higher Berry curvature for gapped 1d systems}\label{sec:one dimension}

As explained in the previous section, given a family of trivial gapped theories on a $D$-dimensional lattice and assuming that the field theory description applies  at each point in the parameter space $\cM$, there should be a way to construct a closed $(D+2)$-form on $\cM$ whose integrals over spherical cycles are quantized. The cohomology class of the form is a topological invariant of the family (cannot change under deformations). In this section we construct such a closed form $\Omega^{(D+2)}$ on $\cM$ for the case of gapped spin chains, that is, gapped lattice $D=1$ systems. We do not use the existence of the field theory limit. In Appendix \ref{appendix:quantization} we argue that integrals of $\Omega^{(D+2)}$ over spherical 3-cycles are quantized. That is, integrals of the form $\int_{S^3}h^*\Omega^{(D+2)}$, where $h$ is a map from $S^3$ to $\cM$, are integer multiples of $2\pi$.

To begin with, let us recall how the Berry 2-form is defined for gapped 0d systems and why this definition does not work for $D>0$. Let $G=1/(z-H)$ be the Green's function for a positive bounded Hamiltonian $H$ which depends on some parameters. Assume that $0$ is an isolated eigenvalue of $H$ for all values of the parameters. Let
\begin{equation}\label{Berry curvature}
\Omega^{(2)}=\frac i {2}\oint \frac{dz}{2\pi i}\Tr (G dH G^2 dH),
\end{equation}
where $\oint$ is the counterclockwise contour integral around $z=0$ and $d$ denotes the exterior derivative on the paramter space $\cM$. That is, $d = \sum_{\ell}^{} d\lambda^\ell \frac{\partial }{\partial \lambda^\ell}$ where $\lambda^\ell$ are parameters. The wedge product of forms $\wedge$  is implicit in Eq. (\ref{Berry curvature}). $\Omega^{(2)}$ is a closed 2-form on $\cM$. Indeed, since $dG=G dH G$, we compute
\begin{multline}
d\Omega^{(2)}=\frac i {2}\oint\frac{dz}{2\pi i} \Tr(GdHGdHG^2dH-GdHG^2dHGdH-GdHGdHG^2dH)=\\
=-\frac i {2}\oint \frac{dz}{2\pi i}\Tr(GdHG^2dHGdH)=
\frac i {6}\oint \frac{dz}{2\pi i}\frac{\partial}{\partial z} \Tr(GdHGdHGdH) =0.
\end{multline}
$\Omega^{(2)}$ is the usual Berry curvature, as one can verify by inserting a complete set of states.

Suppose now $H$ is a many-body Hamiltonian for an infinite 1d lattice system with an energy gap. More explicitly, we assume that  $H=\sum_{p\in \Lambda} H_p$ where $H_p$ is bounded and finite-range and $\Lambda\subset\RR$ is a discrete subset of real numbers without accumulation points. Then $H$ is unbounded, but one can still define a bounded operator $G=1/(z-H)$ for $z$ which are away from the spectrum of $H$. We assume again that $H$ is positive and that $0$ is an isolated eigenvalue for all values of the parameters.  Fixing $p,q\in\Lambda$, we can define
a non-closed 2-form on the parameter space 
$$
\Omega^{(2)}_{pq}=\frac i {2}\oint \frac{dz}{2\pi i}\Tr(G dH_p G^2 dH_q).
$$
If the Hamiltonian $H$ is gapped, $\Omega^{(2)}_{pq}$ decays exponentially away from $p=q$ (see \cite{Watanabe}). 
The Berry curvature is formally given by  
$$
\Omega^{(2)}=\sum_{p,q\in\Lambda} \Omega^{(2)}_{pq},
$$
but the contribution of the points near the diagonal, $p\simeq q$, is divergent for infinite-volume systems.

Instead of the ill-defined Berry curvature 2-form, consider the following 2-form depending on a site $p$:
\begin{equation}
F^{(2)}_p=\frac i {2}\oint \frac{dz}{2\pi i}\Tr (GdHG^2 dH_p).
\end{equation}
It is well-defined, but not closed. Instead one has an identity
\begin{equation}\label{descendant eq d=1}
dF^{(2)}_q=\sum_{p\in \Lambda} F^{(3)}_{pq},
\end{equation}
where the 3-form $F^{(3)}_{pq}$ is given by 
$$
F^{(3)}_{pq}=\frac i {6}\oint  \frac{dz}{2\pi i}\Tr(G^2dHGdH_p GdH_q-GdHG^2dH_p GdH_q)  -(p\leftrightarrow q).
$$
The identity (\ref{descendant eq d=1}) can be verified by a straightforward computation. Note that $F^{(3)}_{pq}$ decays exponentially away from the diagonal $p=q$ thanks to the results of \cite{Watanabe}.

The identity (\ref{descendant eq d=1}) and other similar identities are key for defining topological invariants of families of gapped systems in one and higher dimensions. In the context of Euclidean lattice systems, analogous identities were first observed by A. Kitaev who used them to define invariants of families of such systems \cite{Kitaev_talk}. In this paper we essentially derive Hamiltonian analogs of Kitaev's formulas. 

 Let $f:\Lambda\ra \RR$ be a function which is $0$ for $p \ll 0$ and $1$ for $p\gg 0$. For example, it could be simply $0$ for $p<a$ and $1$ for $p\geq a$. Then we define a 3-form on the parameter space by
\begin{equation}\label{omega f}
\Omega^{(3)}(f)=\frac 1 2 \sum_{p,q\in \Lambda} F^{(3)}_{pq} (f(q)-f(p)).
\end{equation}
It is well-defined because on the one hand $F^{(3)}_{pq}$ decays exponentially for large $|p-q|$, and on the other hand $f(q)-f(p)$ is non-zero only when $p>a$ and $q<a$, or the other way around.  For the specific choice of $f(p)$ equal $0$ for $p<a$ and $1$ for $p\geq a$, the equation (\ref{omega f}) takes a simple form
\begin{equation}
\Omega^{(3)}(f)= \sum_{p<a \atop q>a} F^{(3)}_{pq},
\end{equation}
which makes its convergence more transparent.

Later in this paper we will show that
\begin{align}\label{descent1d}
    dF^{(3)}_{qr}= \sum_{p\in \Lambda} F^{(4)}_{pqr},
\end{align}
where $F^{(4)}_{pqr}$ is a function which is anti-symmetric in $p,q,r$ and decays exponentially away from the diagonal $p=q=r$. We find
\begin{align}\label{closedness in one dimension}
\begin{split}
    d\Omega^{(3)}(f)= \frac 1 2  \sum_{q,r\in\Lambda}(f(r)-f(q)) dF^{(3)}_{qr}=\frac 1 2  \sum_{p,q,r\in\Lambda}  (f(r)-f(q)) F^{(4)}_{pqr}\\=\frac 1 6 \sum_{p,q,r\in\Lambda}  (f(r)-f(q)+f(p)-f(r)+f(q)-f(p)) F^{(4)}_{pqr}=0,
\end{split}
\end{align}
where we have used the anti-symmetry of $F^{(4)}_{pqr}$. Therefore the 3-form $\Omega^{(3)}(f)$ is closed.

Closedness of $\Omega^{(3)}(f)$ implies that its cohomology class is a topological invariant of the family of gapped systems. Indeed, let us regard $\cM$ as a submanifold in the space $\fM_D$ of all gapped systems in dimension $D$. Obviously, the form $\Omega^{(D+2)}$ is a restriction of a closed form on $\fM_D$ defined in exactly the same way. Deforming $\cM$ within $\fM_D$ can be thought of as a flow along a vector field on $\fM_D$. Since the Lie derivative of a closed form along any vector field is exact, deforming $\cM$ cannot change the cohomology class of $\Omega^{(D+2)}$.

The cohomology class of the 3-form $\Omega^{(3)}(f)$ is independent of the choice of the function $f$ as long as $f(p)=0$ for $p\ll 0$ and $f(p)=1$ for $p\gg 0$. Indeed, any two such functions differ by a function $g$ which is compactly supported, and for such a function we can write 

\begin{align}\label{independence one dimension}
    \begin{split}
\Omega^{(3)}(g)=\frac 1 2 \sum_{p,q\in \Lambda} (g(q)-g(p))\Fth_{pq}= \sum_{q\in \Lambda} g(q) \sum_{p \in \Lambda} F^{(3)}_{pq} = \sum_{q\in \Lambda} g(q) d F^{(2)}_q=d\sum_{q\in \Lambda} g(q)  F^{(2)}_q.
    \end{split}
\end{align}
This means that $\Omega^{(3)}(f+g)$ and $\Omega^{(3)}(f)$ differ by a total derivative of a well-defined 2-form on $\cM$ and therefore are in the same cohomology class.

We note the following obvious properties of the 3-form  $\Omega^{(3)}(f)$. It vanishes for constant families (i.e. families where the Hamiltonian is independent of parameters), and it is additive under stacking of families (with the same parameter space).

\section{Higher Berry curvature for gapped systems in any dimension}\label{sec:arbitrary dimension}

To construct analogs of Berry curvature in higher dimensions, the language of chains and cochains is very useful. Let $\Lambda$ be a discrete subset of $\mathbb R^D$ without accumulation points. For $n\ge0$, an $n$-chain is a quantity $A_{p_0 \dots p_n}$ which depends on $n+1$ points $p_0,\ldots,p_n \in\Lambda$, is skew-symmetric under permutations of $p_0,\ldots,p_n$, and decays exponentially away from the diagonal $p_0=p_1=\ldots=p_n$. 
The space of $n$-chains will be denoted $\C_n(\Lambda)$. The boundary operator $\partial:\C_n(\Lambda)\ra \C_{n-1}(\Lambda)$ is defined as follows:
$$
(\partial A)_{p_1 \dots p_n}=\sum_{p_0\in\Lambda} A_{p_0 \dots p_n}.
$$
It is easy to see that $\partial^2=0.$ Thus $\oplus_{n\geq 0}\, \C_n(\Lambda)$ is a chain complex.

Dually, an $n$-cochain (with values in reals) is a real-valued function $\alpha(p_0,\ldots,p_n)$ which depends on $p_0,\ldots,p_n\in\Lambda$, is bounded, skew-symmetric under permutations, and obeys the following
condition: when restricted to any $\delta$-neighborhood of the diagonal, it vanishes when any of the points is outside some finite set. Let $\C^n(\Lambda)$ be the space of $n$-cochains. There is a pairing between $\C^n(\Lambda)$ and $\C_n(\Lambda)$ defined by
\begin{equation}
\langle A,\alpha\rangle=\frac{1}{(n+1)!}\sum_{p_0,\ldots,p_n} A_{p_0,\ldots , p_n}\alpha(p_0,\ldots,p_n)
\end{equation}
There is also an operator $\delta: \C^n(\Lambda)\ra \C^{n+1}(\Lambda)$ satisfying $\delta^2=0$ and uniquely defined by the condition  
\begin{equation}\label{stokes}
\langle A, \delta\alpha\rangle=\langle \partial A,\alpha\rangle
\end{equation}
for any $(n+1)$-chain $A$ and an $n$-cochain $\alpha$ and $n\geq 0$. One can regard (\ref{stokes}) as a version of Stokes' theorem. Explicitly, the operator $\delta$ is given by
\begin{equation}
(\delta \alpha)(p_0,\ldots,p_{n+1})=\sum_{j=0}^{n+1} (-1)^j \alpha(p_0,\ldots,p_{j-1},p_{j+1},\ldots,p_{n+1}).
\end{equation}
In particular, if $\Lambda\subset\RR$ is a 1d lattice, and $f:\Lambda\ra\RR$ is a function such that $f(p)=1$ for $p\gg 0$ and $f(p)=0$ for $p\ll 0$, then $(\delta f)(p,q)=f(q)-f(p)$ is a closed 1-cochain on $\Lambda$. It is not exact, since $f$ does not have a finite support. 

One can define the product $\alpha\cup\gamma$ of an $n$-cochain $\alpha$ and an $m$-cochain $\gamma$ as an $n+m$-cochain given by
\begin{align*}
(\alpha\cup\gamma)(p_0,\ldots,p_{n+m})=\frac{1}{(n+m+1)!}\sum_{\sigma\in {\mathcal S}_{n+m+1}} (-1)^{{\rm sgn}\, \sigma} \alpha(p_{\sigma(0)},\ldots,p_{\sigma(n)})\gamma(p_{\sigma(n)},\ldots,p_{\sigma(n+m)}),
\end{align*}
It satisfies
\begin{equation}
\alpha\cup\gamma=(-1)^{nm}\gamma\cup\alpha,\quad \delta(\alpha\cup\gamma)=\delta\alpha\cup\gamma+(-1)^n\alpha\cup\delta\gamma.
\end{equation}

Using this notation, we see that $\Omega^{(3)}(f)=\langle F^{(3)},\delta f\rangle$, where $F^{(3)}$ is a 1-chain with values in 3-forms on $\cM$ with components $F^{(3)}_{pq}.$ Furthermore, eq. (\ref{descent1d}) can be written as a relation between a 1-chain $F^{(3)}$ valued in 3-forms and a 2-chain $F^{(4)}$ valued in 4-forms:
\begin{equation}\label{descent1dchain}
dF^{(3)}=\partial F^{(4)}.
\end{equation}
Then the computation leading to (\ref{closedness in one dimension})  can be shortened to
$$
d\Omega^{(3)}(f)=\langle d \Fth,\delta f\rangle=\langle\partial F^{(4)},\delta f\rangle=\langle F^{(4)},\delta \delta f\rangle=0.
$$
Similarly, the computation leading to (\ref{independence one dimension}) can be shortened to
$$
\Omega^{(3)}(g)=\langle \Fth,\delta g\rangle =\langle \partial\Fth, g\rangle=d\langle F^{(2)}, g\rangle.
$$
Here $g:\Lambda\ra\RR$ is supported on a finite set, therefore the application of the Stokes' theorem is legitimate.

Now we will generalize the construction of the previous section to arbitrary dimensions and define a closed $(D+2)$-form $\Omega^{(D+2)}$ on the parameter space of a family of gapped lattice systems in $D$ spatial dimensions. We define a family of $D$-dimensional gapped lattice systems in the same way as for $D=1$, the  only difference being that the lattice $\Lambda$ is a subset of $\mathbb R^D$ instead of $\mathbb R$.  For $D>1$ not all gapped systems are can be continuously connected to the trivial one, thanks thanks to the possibility of topological order. Therefore we do not expect our $(D+2)$-form to have quantized periods even on spherical cycles. Nevertheless we will argue in Appendix A that for families of systems in an SRE phase its periods are quantized on spherical $(D+2)$-cycles, as expected from the field theory analysis.

We will define higher Berry curvatures recurrently via the following "descent equation":
\begin{equation}\label{descent equation}
    dF^{(n)}=\partial F^{(n+1)},
\end{equation}
where $F^{(n)}$ is $(n-2)$-chain with values in $n$-forms on the parameter space. Analogous equations for families of Euclidean lattice systems were used in \cite{Kitaev_talk}. Starting from $F^{(2)}$ defined in (\ref{descendant eq d=1}), we can find all its descendants. The result is
\begin{align}
\begin{split}
    F^{(n)}_{p_0 \dots p_{n-2}} &= \frac{i (-1)^n}{ n(n-1)} \sum_{\sigma \in \mathcal S_{n-1}}{ {\rm sgn}( \sigma)}  \oint \frac{dz}{2\pi i} \\ &\sum_{j=0}^{n-2} (n-j-1) \Tr\left( GdH G dH_{p_{\sigma(0)}}GdH_{p_{\sigma(1)}}\dots G^2 dH_{p_{\sigma(j)}} \dots G dH_{p_{\sigma(n-2)}} \right).
\end{split}
\end{align}
For this to be a well-defined chain, it must decay exponentially when any two of the points $p_0,\ldots,p_{n-2}$ are separated by a large distance. For $n=3$ this was proved in \cite{Watanabe}, and we expect that the proof can be generalized to arbitrary $n$. Heuristically, exponential decay follows from the physical interpretation of the above correlators in terms of generalized local susceptibilities. For $n=2$ the correlator is a variation of the expectation value of a local operator $dH_{p_0}$ with respect to an arbitrary infinitesimal variation of the Hamiltonian. That is, it is a local susceptibility. For $n=3$ it can be interpreted as a variation of a local susceptibility with respect to a variation of the Hamiltonian elsewhere. For $n=4$ it can be interpreted as a variation of a variation, etc. We expect all such quantities to decay exponentially for large spatial separations because the correlation length is finite for a gapped system at zero temperature. 

In order to find a topological invariant of a family of gapped systems we need to contract this $(n-2)$-chain with an $(n-2)$-cochain. Let $\alpha$ be an  $(n-2)$-cochain, then $\langle F^{(n)},\alpha\rangle$ is an $n$-form on the parameter space. But in general it is not closed:
\begin{align}
    d\langle F^{(n)},\alpha\rangle = \langle dF^{(n)},\alpha\rangle= \langle \partial F^{(n+1)}, \alpha\rangle=\langle F^{(n+1)}, \delta \alpha\rangle.
\end{align}
In order for the integral of the $n$-form $\int_{C_n}\langle F^{(n)},\alpha\rangle$ to be independent of the deformation of the cycle $C_n$, the cochain $\alpha$ must be closed,  $\delta\alpha=0.$ On the other hand, if the cochain $\alpha$ is exact, $\alpha=\delta\gamma$, we find
\begin{equation}
    \langle F^{(n)},\alpha\rangle = \langle F^{(n)},\delta \gamma\rangle = \langle \partial F^{(n)},\gamma\rangle = d\langle F^{(n-1)},\gamma\rangle ,
\end{equation}
and all integrals $\int_{C_n}\langle F^{(n)},\alpha\rangle$ over cycles $C_n$ will be zero.

We see that in order to get a non-trivial invariant of a family we need to contract the chain $F^{(n)}$ with a cochain which is closed but not exact. Moreover, adding to such a cochain an exact cochain will not change the invariant. Thus we need to understand the space of closed cochains modulo the subspace of exact cochains, that is, the cohomology of the cochain complex $(\C^n(\Lambda),\delta)$. If we omit the word "bounded" from the definition of cochains, then the cohomology of the corresponding complex is known in the mathematical literature as the coarse cohomology of $\Lambda$ \cite{Roe}. For physical applications, one may assume that $\Lambda\subset \mathbb R^D$ uniformly fills the whole $\RR^D$, in the sense that there exists $\delta>0$ such that each point of $\RR^D$ is within distance $\delta$ of some point of $\Lambda$, and that $\Lambda$ has no accumulation points. Then the $n$-th coarse cohomology group of $\Lambda$ is isomorphic to the $n$-th  cohomology group of $\RR^D$ with compact support \cite{Roe}. The latter is non-trivial only for $n=D$ and is one-dimensional. The generator of $D$-th coarse cohomology group can be taken to be  $\delta f_1\cup \dots\cup \delta f_D$, where $f_\mu(p)=\theta(x^\mu(p))$ and $x^\mu(p)$ is the $\mu$-coordinate of $p$ and $\theta(x)$ is theta function. More generally, one can choose $f_\mu$ to be any function which depends only on $x^\mu(p)$ and is $0$ for $x^\mu (p) \ll 0$ and $1$ for $x^\mu(p)\gg 0$. Note that such cochains are bounded and thus also define a nontrivial cohomology class in the sense that we need. For a family of $D$-dimensional systems parameterized by $\cM$ we therefore define a $(D+2)$-form on $\cM$:
\begin{align}
    \Omega^{(D+2)}(f_1,\dots,f_D)= \langle F^{(D+2)}, \delta f_1\cup \dots\cup \delta f_D\rangle .
\end{align}

This $(D+2)$-form is closed:
\begin{align*}
     d\Omega^{(D+2)}(f_1,\dots,f_D)=\langle dF^{(D+2)}, \delta f_1\cup \dots\cup \delta f_D\rangle=\langle \partial F^{(D+3)}, \delta f_1\cup \dots\cup \delta f_D\rangle\\=\langle F^{(D+3)},\delta( \delta f_1\cup \dots\cup \delta f_D)\rangle=0.
\end{align*}
Its cohomology class is unchanged under the shift $f_1\rightarrow f_1+g$ by a compactly-supported function $g$ since
\begin{align*}
    \Omega^{(D+2)}(g,f_2,\dots,f_D)= \langle F^{(D+2)}, \delta g\cup\delta f_2\cup \dots\cup \delta f_D\rangle=\langle F^{(D+2)}, \delta (g\cup\delta f_2\cup \dots\cup \delta f_D)\rangle\\=\langle \partial F^{(D+2)},  g\cup\delta f_2\cup \dots\cup \delta f_D\rangle=d\langle  F^{(D+1)},  g\cup\delta f_2\cup \dots\cup \delta f_D\rangle
\end{align*}
and analogously for other shifts $f_\mu\rightarrow f_\mu+g$. 

In general, periods of $\Omega^{(D+2)}(f_1,\ldots,f_D)$ are not subject to quantization.  In Appendix \ref{appendix:quantization} we argue that integrals over spherical cycles in the parameter space are quantized for families of SRE systems.

\section{Discussion}\label{sec:discussion}
 In this paper, we have constructed higher-dimensional generalizations of the Berry curvature starting from the ordinary Berry curvature for quantum-mechanical systems and solving the descent equation (\ref{descent equation}) . In fact, this procedure of constructing higher-dimensional generalization of topological invariants from lower dimensional ones via descent equations is rather general. For example, the Thouless charge pump for 1d systems \cite{Thouless} and its higher-dimensional generalizations can be constructed from the ground-state charge of a quantum-mechanical system with a $U(1)$ symmetry. This will be discussed in a separate publication.
 
 The cohomology class of the Berry curvature is an obstruction to having a continuously varying family of ground states. Similarly one can show that the cohomology class of the form $\Omega^{(D+2)}$ restricted to a $(D+2)$-sphere in the parameter space is an obstruction to having a smoothly varying gapped boundary defined everywhere on this sphere. See Appendix A for details. An example of a three-parameter family of 1d lattice systems whose cohomology class is non-trivial is given in Appendix \ref{free fermions}, where we relate the 3-form $\Omega^{(3)}(f)$ to the Berry-Bloch connection over the Brillouin zone.
 
 For $D=0$ the cohomology class of the Berry curvature (regarded as an integral class) is the only topological invariant of the family. It is trivial if and only if the family can be deformed to a constant family without closing the gap. One can ask if the same is true for $D>0$ or if there are additional independent invariants. The existence of topological order for $D>1$ means that the answer will probably depend on which topological phase one considers. The case $D=1$ is special since all gapped 1d systems are Short-Range Entangled. Moreover, for $D=1$ it has been conjectured by A. Kitaev that a properly defined space of all gapped bosonic systems has the homotopy type $K(\ZZ,3)$. That is, its only non-trivial homotopy group is in degree $3$ and is isomorphic to $\ZZ$. If this is true, then all cohomology classes on the space of gapped bosonic 1d systems can be expressed as some complicated functions of the basic class which sits in degree $3$. That is, for $D=1$ bosonic families there are no further independent invariants beyond the one we constructed.

\appendix
\section{Quantization of higher Berry curvatures} \label{appendix:quantization}
Consider a family of gapped systems in spatial dimension $D$. In the body of the paper we showed how to define a closed form $\Omega^{(D+2)}$ on the parameter space $\cM$. It depends on some additional data ($D$ functions on $\Lambda$), but the cohomology class was shown to be independent of these data. Thus periods of $\Omega^{(D+2)}$ are also independent of these additional data. In this appendix we argue that if all systems in the family are Short-Range Entangled (SRE), and if $h$ is a spherical cycle in $\cM$ (i.e. a map  $h:S^{D+2}\ra \cM$), then the integral of $\Omega^{(D+2)}$ over such a cycle is "quantized":
\begin{equation}
\frac 1 {2\pi}\int_{S^{D+2}} h^* (\Omega^{(D+2)})\in \ZZ.
\end{equation}

We begin with the 1d case, where there is no topological order, and thus all gapped systems without spontaneous symmetry breaking are SRE. Thus all systems in the family belong to the same SRE phase. In the bosonic case, this means that they can all be deformed to a trivial system whose Hamiltonian is a sum of one-site operators and the ground state is a product state. In the fermionic case, there is a unique non-trivial SRE phase corresponding to Kitaev's Majorana chain. So there are two options: either all systems in the family are in the trivial phase, or they can all be deformed to the Majorana chain. In the latter case we can stack the whole family with the "constant" Majorana chain and get a family of fermionic systems in the trivial phase.  Since $\Omega^3(f)$ is unchanged under stacking the family 
with a system independent of parameters, this reduces the problem to studying a family of systems in the trivial phase.

Let $f(p)=\theta(p)$ (a step-function on $\Lambda\subset\RR$). Recall that we denote the space of all gapped 1d system by $\fM_1$. (Our argument will be the same for bosonic and fermionic systems, so we do not need to distinguish the two possibilities). This is an infinite-dimensional space which can be thought of as a union of an infinite number of finite-dimensional manifolds. The parameter space $\cM$ is a submanifold in this infinite-dimensional space, and the 3-form $\Omega^{(3)}$ on $\cM$ is a restriction of the 3-form on $\fM_1$ defined in exactly the same way. Let us fix a particular trivial system $\cm_0\in\fM_1$. Each point in $\cM$ can be connected to $\cm_0$ by a continuous path in $\fM_1$. This applies to all points in the image of the spherical cycle $h$. If this could be done continuously over the whole $S^3$, it would mean that the cycle is contractible to a point $\cm_0$ in $\fM_1$, and the corresponding integral $\int_{S^3}h^*(\Omega^{(3)}(f))$ would be zero. While in general it is not possible to contract the whole spherical cycle, it is always possible to contract $S^3$ with a point removed. In particular, it is possible to contract $S^3$ without either north or south pole. Let $S^3_S$ and $S^3_N$ be $S^3$ with the north and south poles removed, respectively. Let us denote the contractions in the space of the gapped Hamiltonians by $\mathcal P_S$ and $\mathcal P_N$. These are continuous maps from $[0,1]\times S^3_S$ to $\fM_1$ and from $[0,1]\times S^3_N$ to $\fM_1$, respectively. Let us parameterize $[0,1]$ by $t$. For $t=0$ they are just restrictions of $h$ to $S^3_S$ and $S^3_N$. For $t=1$ they are constant maps to $\cm_0$.

Let the Hamiltonian corresponding to a point $\cm\in \fM_1$ be $H(\cm)=\sum_p H_p(\cm)$. The family of Hamiltonians corresponding to the spherical cycle $h$ is $H[\cs]=\sum_p H_p(h(\cs))$, where $\cs\in S^3$. 
For $\cs\in S^3_N$  we define another Hamiltonian $H^+[\cs]$ which is the same as $H[\cs]$ except that on the far right part of the lattice $p\gg 0$ it adiabatically interpolates to $H(\cm_0)$. More precisely, $H^+[\cs]=\sum_{p\in\Lambda} H^+_p[\cs]$ is sum of on-site Hamiltonians $H^+_p[\cs] = H_p(\cm(\cs,p))$ where we let the parameters of the Hamiltonian depend slowly on $p$ as $\cm(\cs,p)=\mathcal P_N(t_N(p),\cs)$. The function $t_N:\RR\ra\RR$ is equal to 1 for $p\in [2L,+\infty)$, smoothly interpolates from 1 to 0 in the region $p\in [L,2L]$, and is 0 for $p\in (-\infty,L]$. Similarly, we define a local Hamiltonian $H^-[\cs]$ for all $\cs\in S^3_S$ via $H^-[\cs]=\sum_{p\in\Lambda} H_p(P_S(t_S(p),\cs))$ where the function $t_S:\RR\ra\RR$ is 1 for $p\in (-\infty,-2L],$ smoothly interpolates from 1 to 0 in the region $p\in [-2L,-L],$ and is 0 for $p\in [-L,+\infty)$. Lastly, we define $H^{+-}_p[\cs]$ for all $\cs\in S^3_N\bigcap S^3_S$ as a Hamiltonian which coincides with $H_p[\cs]$ in the  region $p\in [-L,L]$, coincides with $H_p(\cm_0)$ for $p\notin [-2L,2L]$, and smoothly interpolates between these regions using the paths $\mathcal P_S$ and $\mathcal P_N$. Our main assumption is that all these families of Hamiltonains are gapped for sufficiently large $L$. This seems reasonable since for a fixed $t$ and $\cs$ all Hamiltonians $H(P_N(t,\cs))$ and $H(P_S(t,\cs))$ are gapped and there should be an upper bound on the correlation length. However, a proof of this would be very desirable.  We denote by $\Omega^{(3)}_+(f),\Omega^{(3)}_-(f)$ and $\Omega^{(3)}_{+-}(f)$ the 3-forms corresponding to the families $H^+$, $H^-$ and $H^{+-}$. They are defined on $S^3_N$, $S^3_S$ and $S^3_N\bigcap S^3_S$, respectively.

We write an integral over $S^3$ as a sum of integrals over its lower and upper hemispheres which we call $B_-$ and $B_+$:
\begin{equation*}
    \int_{S^3} h^*(\Omega^{(3)}(f) )= \int_{B_+} h^*(\Omega^{(3)}(f))+ \int_{B_-} h^*(\Omega^{(3)}(f)) =\int_{B_+} \Omega^{(3)}_+(f)  +\int_{B_-} \Omega^{(3)}_-(f) + O(L^{-\infty}).
\end{equation*}
In the last step we replaced $h^*(\Omega^{(3)})$ with $\Omega^{(3)}_\pm$ on $B_\pm$. Since by our assumption $H[\cs]$, $H^+[\cs],$ and $H^-[\cs]$ are all gapped, the 3-form $h^*(\Omega^{(3)})$ is only sensitive to the Hamiltonian of the system in the neighborhood of the point $p=0$ where the function $f(p)=\theta(p)$ has a discontinuity. Since all these Hamiltonians coincide near the point $p=0$, for large $L$ the error introduced by this replacement is of order $L^{-\infty}$.

Let us now define $f_+(p)=\theta(p-3L)$ and $f_-(p)= \theta(p+3L)$ and write 
\begin{equation}
 \int_{B_+} \Omega^{(3)}_+(f)  +\int_{B_-} \Omega^{(3)}_-(f)=\int_{B_+} \Omega^{(3)}_+(f_+)  +\int_{B_-} \Omega^{(3)}_-(f_-)+\int_{B_+} \Omega^{(3)}_+(f-f_+)  +\int_{B_-} \Omega^{(3)}_-(f-f_-),
\end{equation}
The on-site Hamiltonian $H_p^+[\cs]$ coincides with the constant Hamiltonian  $H_p(\cm_0)$ near $p=3L$. Therefore the form $\Omega^{(3)}_+(f_+)$ is of order $L^{-\infty}$, and so is its integral over $B_+$. Similarly, $\int_{B_-} \Omega^{(3)}_-(f_-) = O(L^{-\infty})$.  The remaining terms in the above equation contain functions $f_\pm -f$ which have compact support. For any such function $g:\Lambda\ra\RR$ we can write $\Omega^{(3)}_\pm(g) =\langle F^{(3)}_\pm,\delta g\rangle = d \langle F^{(2)}_\pm, g\rangle $. Therefore we get
\begin{equation}
\int_{B_+} \Omega^{(3)}_+(f-f_+)  +\int_{B_-} \Omega^{(3)}_-(f-f_-) = \int_{S^2} \langle F^{(2)}_+,f-f_+\rangle  -\int_{S^2} \langle F^{(2)}_-,f-f_-\rangle
\end{equation}
where $S^2$ is the equator of $S^3$ and the common boundary of $B_-$ and $B_+$. The minus sign arises because the orientation on $S^2$ induced by $B_-$ is opposite to the one induced by $B_+$. We can now replace $F^{(2)}_+$ and $F^{(2)}_-$ with $F^{(2)}_{+-}$ in both integrals, since the integrands are only sensitive to the Hamiltonian of the system in the region where $H^+_p[\cs]=H^{+-}_p[\cs]$ and $H^-_p[\cs]=H^{+-}_p[\cs]$. Such a replacement introduces an error of order $L^{-\infty}$. Therefore the above expression becomes
\begin{align}\begin{split}
\int_{S^2} \langle F^{(2)}_+,f-f_+\rangle  -\int_{S^2} \langle F^{(2)}_-,f-f_-\rangle = \int_{S^2} \langle F^{(2)}_{+-}, f-f_+\rangle  -\int_{S^2} \langle F^{(2)}_{+-}, f-f_-\rangle+O(L^{-\infty})  \\= - \int_{S^2} \langle F^{(2)}_{+-},f_+-f_-\rangle  +O(L^{-\infty}). 
\end{split}
\end{align}
By construction $H^{+-}_p[\cs]=H_p[\cs]$ for $p\in [-L,L]$, while $H^{+-}_p[\cs]=H(\cm_0)$ for $p\notin [-2L,2L]$.  Since outside $[-2L,2L]$ the Hamiltonian is constant, that part of the system does not contribute to $F^{(2)}$ and can be discarded. What remains is a system with a finite-dimensional Hilbert space. Since $f_+-f_-= \theta(p-3L)-\theta(p+3L)$ and thus is equal $-1$ in the region $[-2L,2L]$, we have 
\begin{equation}
-\langle F^{(2)}_{+-},f_+-f_-\rangle=\sum_{p\in [-2L,2L]} F^{(2)}_{+- p}+O(L^{-\infty}).
\end{equation}
This is simply the Berry curvature of this finite-dimensional system. Therefore its integral over $S^2$ is an integer multiple of $2\pi$. We conclude that 
\begin{equation}
\int_{S^3} h^*(\Omega^{(3)}(f))=2\pi n+O(L^{-\infty}), \quad n\in\ZZ.
\end{equation}
Taking the limit $L\ra\infty$ we get the desired result.

In general we proceed by induction in $D$. For $D>1$ the restriction to SRE systems is a nontrivial constraint on the kind of families we allow. Other than that, we can proceed in the same way as for $D=1$. First we tensor with a suitable constant SRE system to reduce to the case of a family of systems in a trivial phase. Then we remove the north and south pole from $S^{D+2}$ and define three families of gapped Hamiltonians $H^+[\cs]$, $H^-[\cs],$ and $H^{+-}[\cs]$ which are defined on $S^{D+2}_N$, $S^{D+2}_S$ and $S^{D+2}_N\bigcap S^{D+2}_S$, respectively. They approach $H(\cm_0)$ on the far right, far left, and both far right and far left, respectively. By far right we mean the region $x^D(p)\gg 0$, while far left is the region $x^D(p)\ll 0$. The same manipulations as before reduce the integral of $\Omega^{(D+2)}$ over $S^{D+2}$ to an integral of $\Omega^{(D+1)}$ over the equatorial $S^{D+1}$ up to terms of order $L^{-\infty}$. This completes the inductive step. 

An interpolation between $H(\cm)$ and $H(\cm_0)$ can also be viewed as a gapped boundary condition for $H(\cm)$. Given a smooth family of gapped  boundary conditions for $H[\cs]$ defined on some open subset $U\subset S^3$ (not necessarily arising from a smooth interpolation as above), one can write $\Omega^{(D+2)}(f_1,\ldots,f_D)\vert_U$ as an exact form. This is done in exactly the same way as above. Therefore if the cohomology class of $\Omega^{(D+2)}$ is non-trivial, it is impossible to find a family of gapped boundary conditions for $H[\cs]$ which is defined on the whole $S^3$ and varies smoothly with $\cs$. For $D=0$ the analogous statement is that the cohomology class of the Berry curvature is an obstruction to finding a family of ground states on the whole parameter space which depends continuously on the parameters.

\section{Higher Berry curvature for 1d insulators of class A} \label{free fermions}

In this appendix we compute the higher Berry curvature 3-form in the case of gapped systems of free fermions in 1d with conserved charge (that is, insulators of class A). Then we specialize to the case of translationally-invariant systems and compare with forms constructed out of the Bloch-Berry connection.

We start with the many-body expression for the 3-form $\Fth_{pq}$ divided by $2\pi$:
\begin{equation}\label{many body f3}
\frac {\Fth_{pq}}{2\pi}=-\frac{i}{12\pi}\oint_{z=E_0} \frac{dz}{2\pi i} \Tr(2GdHG^2 dH_p G dH_q+GdHGdH_p G^2dH_q) -(p\leftrightarrow q)
\end{equation}

We will consider the following many-body Hamiltonian:
\begin{equation}
H_p=\frac12\sum_{m\in\Lambda} \left(a^\dagger_ph(p,m)a_m+a^\dagger_m h(m,p) a_p\right).
\end{equation}
Here $h(p,q)$ is an Hermitian matrix $h(p,q)^*=h(q,p)$. The fermionic creation-annihilation operators $a_p^\dagger, a_p$ satisfy canonical anti-commutation relations 
\begin{align}
    \begin{split}
          \{ a^\dagger_p, a_q\}&= \delta_{pq},\\
            \{ a_p, a_q\}&=      \{ a^\dagger_p, a^\dagger_q\}=0,
      \end{split}
\end{align}
where $\delta_{pq}$ is the Kronecker delta.

Since all relevant operators are sums of single particle operators, matrix elements $\langle m |A| n\rangle $ vanish unless many-body states $n$ and $m$ differ by exactly one single-particle excitation. The above expression can be written in terms of one-particle quantities as follows:
\begin{align}
    \frac {\Fth_{pq}}{2\pi}=-\frac{i}{12\pi}\oint \frac{dz}{2\pi i}  {\rm tr}(2gdhg^2 dh_p g dh_q+gdhgdh_p g^2dh_q)  -(p\leftrightarrow q).
\end{align}
Here the contour of integration encloses all states below Fermi level and all lower case letters denote the corresponding single-particle  operators acting on the single-particle Hilbert space $\ell^2(\Lambda)$. Naively, this integral contains additional contributions compared to (\ref{many body f3}) where a fermion jumps from an empty state or jumps into a filled state. But these contributions cancel each other and the result coincides with (\ref{many body f3}).

Hamiltonian density at a point $p$ can be written as $h_p= \frac  1 2 (\delta_p h + h\delta_p)$, where $\delta_p$ is Kronecker's delta (equal 1 on $p$ and 0 on other cites) and functions are understood as operators on the one-particle Hilbert space acting by multiplication. Contracting $\Fth_{pq}$ with the cochain $f(q)-f(p)$ we find
\begin{align}\label{F3 df}
\begin{split}
   \frac{1}{2\pi}\langle \Fth,\delta f\rangle = -\frac{i}{24\pi}\oint\frac{dz}{2\pi i} {\rm tr}\Big (&[dh,f](gdhgdhg^2-g^2dhgdhg)\\&-2[h,f](gdhg^2dhgdhg-gdhgdhg^2dhg) \Big).
\end{split}
\end{align}
Note that multiplication by $f$ is not a trace class operator, since it acts on infinitely many sites. Therefore traces containing them are not guaranteed to exist. On the other hand, commutators like $[dh,f]$ are supported only on a finite number of sites and traces containing them are well-defined. 

On the other hand, given a gapped 1d system of free fermions with translational symmetry which depends on three parameters $\lambda_1,\lambda_2,\lambda_3$, one may consider the Bloch bundle of filled states over the product of the Brillouin zone $S^1$ and the parameter space $\Sigma$. It carries the non-Abelian Bloch-Berry connection, and one can consider various Chern-Weil forms on $S^1\times\Sigma$ constructed out of this connection. In particular, one can consider the degree-4 component of the Chern character of the Berry-Bloch connection and its integral over $S^1\times\Sigma$:
\begin{equation}
\int_{S^1\times\Sigma} {\rm Ch}(\cf)=-\frac 1 {8\pi^2}\int_{S^1\times\Sigma} \Tr(\cf\wedge \cf).
\end{equation}
Here $\cf$ is the non-Abelian curvature 2-form of the Bloch-Berry connection and trace is taken over filled bands. 
It can be shown (see Sec. IIIA in \cite{Qi Hughes Zhang}) that this quantity can be expressed in terms of the one-particle Green's function as follows:
\begin{align}
\begin{split}
   &-\frac 1 {8\pi^2}\int_{S^1\times\Sigma} \Tr(\cf\wedge \cf)\\ &= \frac{\pi^2}{15} \epsilon^{\mu\nu\rho\sigma\tau} \oint \frac{dz}{2\pi i} \int_{S_1} \frac{dk}{2\pi}\int_{\Sigma} \frac{d^3\lambda}{(2\pi)^3} {\rm tr'}\left[ \left(g \frac{\partial g^{-1}}{\partial q^\mu}\right)\left(g \frac{\partial g^{-1}}{\partial q^\nu}\right)\left(g \frac{\partial g^{-1}}{\partial q^\rho}\right)\left(g \frac{\partial g^{-1}}{\partial q^\sigma}\right)\left(g \frac{\partial g^{-1}}{\partial q^\tau}\right)\right],
\end{split}
\end{align}
where $q^\mu=(z,k,\lambda_1 ,\lambda_2,\lambda_3)$. The first integral encloses filled levels, the second integral is over the Brillouin zone, and the last integral is over the parameter space $\Sigma$. The trace ${\rm tr'}$ is taken over subspace with fixed momentum $k$. In translationally invariant system we can interpret $ \int_{S_1} \frac{dk}{2\pi}$ as part of the trace ${\rm tr}$ over the whole one-particle Hilbert space and substitute $\frac{\partial g^{-1}}{\partial k} = -\frac{\partial h}{\partial k} =- i[h,f]$. Expanding the derivatives $\partial/\partial q^\mu$ and combining parameter derivatives into forms, $\sum_i\dfrac{\partial h}{\partial \lambda^i}d\lambda^i = dh $, we find
\begin{multline}
   -\frac 1 {8\pi^2}\int_{S^1\times\Sigma} \Tr(\cf\wedge \cf)=\frac{i}{24\pi} \oint \frac{dz}{2\pi i} \int_{\Sigma} {\rm tr}\Big( g^2[h,f]gdhgdhgdh\\-g^2dhg[h,f]gdhgdh+g^2dhgdhg[h,f]gdh-g^2dhgdhgdhg[h,f]\Big).
\end{multline}
One can see that the integrand of this expression differs from (\ref{F3 df}) by a total derivative proportional to
\begin{align}
    d \left( \oint\frac{dz}{2\pi i} {\rm tr}\Big ([h,f](gdhgdhg^2-g^2dhgdhg) \Big) \right).
\end{align}
Since $\Sigma$ was an arbitrary three-dimensional submanifold of the parameter space, we have shown that the first higher Berry 3-form divided by $2\pi$ is in the same cohomology class as $\int_{S^1\times\Sigma} {\rm Ch}(\cf)$. We conjecture that more generally for class A insulators in $D$ dimensions the form $\Omega^{(D+2)}$ is in the same cohomology class as the integral of the degree $2D+2$ component of the Chern character of the Bloch-Berry connection over the Brillouin zone.

An example of a free 1d fermion system with a non-trivial integral $\int_{S^1\times\Sigma} {\rm Ch}(\cf)$ can be constructed using the 4d Chern insulator (see sec. IIIB of \cite{Qi Hughes Zhang}). The Hamiltonian is 
\begin{align}
    H=\sum_{k_x}\psi^\dagger_{k_x} d_a(k_x,{\vec \lambda}) \Gamma^a \psi_{k_x},
\end{align}
where $\Gamma^a$ are five Dirac matrices generating a Clifford algebra, and 
\begin{equation}
    d_a(k_x,{\vec \lambda }) = \left[(m+c+\cos k_x+c\sum_{i=1}^3 \cos \lambda_i), \sin k_x, \sin \lambda_1,\sin \lambda_2,\sin \lambda_3 \right].
\end{equation}
It was shown in Ref. \cite{Qi Hughes Zhang} that if we chose $\Sigma$ to be 3-torus $S^1\times S^1\times S^1$ defined by identification $\lambda_i\sim \lambda_i +2\pi$ this model has a non-zero integer value of the integral  $\int_{S^1\times S^1\times S^1\times S^1} {\rm Ch}(\cf)$ for a particular choice of $m$ and $c$. One can think about this family of 1d models as a "dimensional reduction" of the 4d Chern insulator where we treat three out of four components of  momentum as parameters. 

Note that the Atiyah-Singer index theorem \cite{AtiyahSinger} implies that the integral of the Chern character of a vector bundle over a four-torus is an integer. Therefore the integral of $\Omega^{(3)}$ over the parameter space $T^3$ is $2\pi$ times an integer, despite the fact that the parameter space is a torus rather than a sphere.

\end{document}